\documentclass[acmlarge,nonacm]{acmart}
\AtBeginDocument{\settopmatter{printacmref=true}}

\usepackage{tabularx}
\usepackage{svg}

 \newcommand\blfootnote[1]{
    \begingroup
    \renewcommand\thefootnote{}\footnote{#1}
    \addtocounter{footnote}{-1}
    \endgroup
}

\copyrightyear{2026}
\acmYear{2026}
\setcopyright{cc}
\setcctype{by}
\acmConference[FAccT '26]{The 2026 ACM Conference on Fairness, Accountability, and Transparency}{June 25--28, 2026}{Montreal, QC, Canada}
\acmBooktitle{The 2026 ACM Conference on Fairness, Accountability, and Transparency (FAccT '26), June 25--28, 2026, Montreal, QC, Canada}
\acmDOI{10.1145/3805689.3806462}
\acmISBN{979-8-4007-2596-8/2026/06}

\begin{document}

\title[The Algorithmic Gaze of Image Quality Assessment]{The Algorithmic Gaze of Image Quality Assessment: An Audit and Trace Ethnography of the LAION-Aesthetics Predictor}

\author{Jordan Taylor}
\orcid{0000-0002-0896-992X}
\email{jordant@andrew.cmu.edu}
 \affiliation{
   \institution{Carnegie Mellon University}
  \city{Pittsburgh}
  \state{PA}
 \country{USA}
 }

\author{William Agnew}
\orcid{0000-0002-1362-554X}
\email{wagnew@andrew.cmu.edu}
 \affiliation{
   \institution{Carnegie Mellon University}
  \city{Pittsburgh}
  \state{PA}
 \country{USA}
 }

\author{Maarten Sap}
\orcid{0000-0002-0701-4654}
\email{msap2@andrew.cmu.edu}
 \affiliation{
   \institution{Carnegie Mellon University}
  \city{Pittsburgh}
  \state{PA}
 \country{USA}
 }

\author{Sarah E. Fox*}
\orcid{0000-0002-7888-2598}
\email{sarahf@andrew.cmu.edu}
 \affiliation{
   \institution{Carnegie Mellon University}
  \city{Pittsburgh}
  \state{PA}
 \country{USA}
 }

\author{Haiyi Zhu*}
\orcid{0000-0001-7271-9100}
\email{haiyiz@andrew.cmu.edu}
 \affiliation{
   \institution{Carnegie Mellon University}
  \city{Pittsburgh}
  \state{PA}
 \country{USA}
 }

\begin{abstract}

Visual generative AI models are trained using a one-size-fits-all measure of aesthetic appeal. However, what is deemed ``aesthetic'' is inextricably linked to personal taste and cultural values, raising the question of \textit{whose} taste is represented in visual generative AI models. In this work, we study an aesthetic evaluation model—the LAION-Aesthetics Predictor (LAP)—that is widely used to curate datasets to train visual generative AI models, like Stable Diffusion, and evaluate the quality of AI-generated images. To understand what LAP measures, we audited the model across three datasets. First, we examined the impact of aesthetic filtering on the LAION-Aesthetics Dataset ($\sim$1.2B images), which was curated from LAION-5B using LAP. We find that the LAP disproportionally filters in images with captions mentioning women, while filtering out images with captions mentioning men or LGBTQ+ people. Then, we used LAP to score $\sim$330k images across two art datasets, finding the model rates realistic images of landscapes, cityscapes, and portraits from western and Japanese artists most highly. In doing so, the algorithmic gaze of this aesthetic evaluation model reinforces the imperial and male gazes found within western art history. In order to understand where these biases may have originated, we performed a trace ethnography of public materials related to the creation of LAP. We find that the development of LAP reflects the biases we found in our audits, such as the aesthetic scores used to train LAP primarily coming from English-speaking photographers and western AI-enthusiasts. In response, we discuss how aesthetic evaluation can perpetuate representational harms and call on AI developers to shift away from prescriptive measures of ``aesthetics'' toward more pluralistic evaluation.

\end{abstract}





\maketitle

\blfootnote{* Co-senior authors contributed equally to this research}

\section{Introduction}

Although artistic taste is deeply subjective and highly political \cite{bourdieu1984distinction}, AI researchers often attempt to quantify the aesthetic quality of media \cite{tjandra2025meta_audio_aesthetic, huang2024aesbench}. While aesthetic quality assessment (AQA) has been a topic of research in the computer vision community for decades \cite{datta2006studying, murray2012ava}, this task is particularly important to examine in light of the rapid growth in visual generative AI. Not only are AQA models used to curate the training data for visual generative AI models \cite{chen2024pixart_sigma}, but they are also used to assess the quality of AI-generated images \cite{gao2024fbsdiff}. As generative AI models are trained on datasets scraped from across the web, researchers rely on computational methods to filter out ``low-quality'' data. Prior work has called attention to myriad gender and cultural biases in common dataset filtering techniques—such as blocklists \cite{dodge2021documenting} and image-text similarity metrics \cite{hong2024datacomp}—but the consequences of aesthetic filtering has yet to be explored in detail. 

In this work, we investigate an AQA model—the LAION-Aesthetics Predictor (LAP)—that has been widely influential in visual generative AI research and commercial development. For example, LAP was used to curate the open-source LAION-Aesthetics Dataset from which Stability AI pre-trained the earliest versions of Stable Diffusion \cite{laion_aesthetics}. Additionally, LAP is often used to \textit{evaluate} the quality of AI-generated images \cite{ding2024understanding_data_poison, laion_eval_1, gao2024fbsdiff, he2023learning}. Despite its far-reaching impact, the LAP itself has yet to be explored in detail.

Due to the subjectivity of aesthetic evaluation, we first explored (RQ1) what the LAP model determines to be a ``high-quality'' image using three datasets. First, we examined the impact of aesthetic filtering on the LAION-Aesthetics Dataset ($\sim$1.2B images), which was curated using LAP. We find that the LAP disproportionally filters in images with captions mentioning women, while filtering out images with captions mentioning men or LGBTQ+ people. Then, we used LAP to score $\sim$330k images across two art datasets, finding the model rates realistic images of landscapes, cityscapes, and portraits from western and Japanese artists most highly. In other words, the \textbf{algorithmic gaze} of LAP reproduces the male gaze within western art history \cite{berger2008ways}. To understand where these biases may have originated, we performed a trace ethnography of public materials to examine (RQ2) how LAP was developed. We find that LAP's development reflects the biases from our audits, such as the aesthetic scores used to train LAP primarily coming from English-speaking photographers and western AI-enthusiasts. 

In light of our findings, we call on AI researchers to contend with aesthetics as a site of cultural representation. Simultaneously, we caution that greater ``inclusion'' in visual generative AI training data risks exacerbating the harms of these technologies. Additionally, we call on developers to shift away from prescriptive, universalist measures of ``aesthetics'' toward more descriptive, pluralistic evaluations. Finally, we discuss how future FAccT researchers might also combine audits and ethnographies of algorithmic systems.

\section{Background \& Related Work}

\subsection{Background on LAION-Aesthetics}

LAION (Large-Scale Artificial Intelligence Open Network) is a non-profit founded in 2021 by Christoph Schuhmann, a German high school teacher, with the goal of creating open-source datasets and machine learning models \cite{laion_bloomberg_2023}. Despite its humble origins, LAION has highly influenced generative image model development. Stability AI researchers used the LAION-Aesthetics Dataset to pre-train the earliest versions of Stable Diffusion \cite{rombach2022stable_diffusion}. Unfortunately, the extent to which \textit{contemporary} generative image models rely on the LAION datasets can be challenging to asses because developers have become cautious of disclosing their training data. While Stability AI shared that early versions of Stable Diffusion were trained using LAION datasets, the model card for their 2024 model Stable Diffusion 3.5-Large vaguely says it was \textit{``trained on a wide variety of data, including synthetic data and filtered publicly available data.''}\footnote{https://huggingface.co/stabilityai/stable-diffusion-3.5-large} This secrecy may be due to the increased scrutiny faced by LAION and generative AI developers \cite{reisner2023revealed, Reisner_2024}, which we will describe below.

LAION has faced substantial criticism from both AI researchers and artists. LAION researchers compiled the 5.86 billion image-text pairs of the LAION-5B dataset by extracting image links and alternative text descriptions from the IMG HTML tags of websites archived by the Common Crawl, a US-based non-profit that has scraped publicly available websites since 2008 \cite{reisner2025common_crawl, baack2024training_common_crawl}. Due to LAION researchers' indiscriminate data collection practices, researchers found that LAION-5B contained over a thousand images of child sexual abuse material (CSAM) \cite{thiel2023identifying}. In addition to containing CSAM, LAION-5B contains millions of copyrighted images, leading to numerous lawsuits against LAION \cite{Robert_Kneschke_v_LAION} and Stability AI \cite{getty_lawsuit}. Furthermore, Birhane et al. found that the image descriptions in LAION-400M and the English language subset of LAION-5B (LAION-2B-en) often contain hate speech \cite{birhane2023laion}.  

Although prior research has examined biases in LAION datasets \cite{birhane2023laion, hong2024datacomp}, the LAION-Aesthetics Dataset and the LAION-Aesthetics Predictor (LAP) model\footnote{https://github.com/christophschuhmann/improved-aesthetic-predictor} used to curate this dataset are underexplored. Not only is the LAION-Aesthetics Dataset widely used to pre-train and fine-tune visual generative AI models, but the LAP model used to to curate this dataset has also been used to curate internal corporate datasets \cite{chen2024pixart_sigma} and to evaluate the performance of visual generative AI models \cite{ding2024understanding_data_poison, laion_eval_1, gao2024fbsdiff, he2023learning}. Although some humanistic research has critiqued how LAP was trained \cite{itavuori2025tate_laion, vianna2025aesthetic}, researchers have not yet empirically investigated what this image quality assessment model measures in practice. In doing so, we contribute both to scholarship on art and AI as well as, more broadly, AI evaluation and alignment practices. In the following sections, we briefly describe each of these contexts in greater detail.

\subsection{AI Evaluation \& Alignment}

AI researchers are increasingly critiquing the measurements used to train and evaluate AI models. For example, benchmarks often claim to measure concepts with no agreed upon definition, such as ``intelligence'' \cite{bean2025measuring}. Researchers have also criticized the notion of universal ``gold standards'' in data annotation \cite{wang2024annotator_disagreement}. While researchers often treat annotator disagreement as noise to be filtered out, disagreements can stem from annotators' differing lived experiences \cite{wang2024annotator_disagreement, sap2022annotators}. In response, researchers have called for distinguishing between datasets intended to represent different beliefs or encode one specific belief \cite{rottger2022two_paradigms}. AI alignment research focuses on normative evaluations where people can reasonably disagree about how a model should behave. For instance, LLMs tend to reflect social values more commonly held in western countries according to national value surveys \cite{tao2024cultural}. At the same time, this methodology can flatten intra-cultural differences \cite{alkhamissi2025hire_anthropologist}. Instead of choosing one set of values to reflect, researchers have increasingly called for designing AI systems more pluralistically \cite{sorensen2024value, sorensen2024roadmap_plural}. In this work, we contend with ``aesthetics'' as an underexamined axis of alignment.

Prior work has found that the design of visual generative AI models often harms marginalized communities. Text-to-image (T2I) models stereotype or poorly represent women \cite{bianchi2023easily}, LGBTQ+ people \cite{ungless2023stereotypes, taylor2025straightening}, people of color \cite{wolfe2022american}, and non-western cultures \cite{fu2024being_eroded, mim2024between, qadri2025non_western_artworld, yerukola2025mindTheGesture}. T2I models made by western tech companies tend to perform worse in languages other than English, such as Farsi \cite{qadri2025non_western_artworld} or Bangla \cite{mim2024between}. This linguistic bias can also be seen in a dataset we examine in this study: the LAION-Aesthetics Dataset is restricted to images with English captions \cite{laion_aesthetics}. HCI researchers have found that artists often enjoy the glitches and strangeness of T2I models \cite{chang2023prompt}, leading some to critique the stylistic bias toward realism in newer models \cite{taylor2025straightening}. In light of this concern, Taylor et al. have called for future research ``examining measurements of aesthetic quality in popular datasets'' \cite{taylor2025straightening}, which we take up in this work.

While the artistic consequences of aesthetic evaluation are underexamined, aesthetic quality assessment (AQA) has been a subject of research in the computer vision community for decades \cite{murray2012ava, datta2006studying}. In this work, researchers formulate AQA as a regression problem predicting a numerical aesthetic score. At the same time, this de-contextualized measurement of aesthetic quality has been critiqued \cite{goree2025human, huang2024aesbench}. \citet{goree2025human} highlight value misalignments between various AQA models and human-preferences by developing an aesthetic evaluation mobile camera application. Like ourselves, \citet{li2025aesbiasbench} recently advocated for investigating aesthetic biases as an underexamined form of representational harms in mulit-modal LLMs. However, the authors conceptualize aesthetic biases as a distributional alignment problem wherein aesthetic evaluation models may privilege the values of one demographic group over others, examining age, gender and education. We extend this work by studying the broader cultural consequences of AQA on data curation and model evaluation, grounding the task within art history.

\section{Auditing the LAION-Aesthetics Predictor Model}
\label{sec:audit}

\subsection{Audit Methodology}

We audited LAP using three datasets. First, we examine the LAION-Aesthetics Dataset curated from LAION-5B by LAP. We then audited LAP using two art datasets with robust metadata on the artists, cultures, styles, and mediums associated with each image. The first dataset consists of 249,351 public-domain images from the Metropolitan Museum of Art (MET).\footnote{https://www.metmuseum.org/hubs/open-access} Located in New York City, the MET is the fourth largest museum in the world and its collection spans over 5,000 years \cite{met_facts}. While the MET dataset contains images from a wide breadth of art across various cultures and time periods, few images are of modern or contemporary art (only 865 images). Therefore, we evaluated LAP on the WikiArt dataset \cite{saleh2015wiki_art} of 81,444 images of visual art by 129 primarily modern (mid-1800s to mid-1900s) artists. WikiArt is a community edited visual art encyclopedia commonly studied in digital humanities research \cite{mohamed2022artelingo_wikiart, fu2025preserving}.

Across each datasets, our analyses primarily focus on understanding differences between images rated above and below an aesthetics score of 6.5. We chose this threshold because it is commonly used in computer vision research to represent high-quality images. The LAION-Aesthetics Dataset was released in multiple subsets based on different aesthetic thresholds, with images scored 4.5+ ($\sim$ 1.2B images) being the lowest rated dataset and 6.5+ ($\sim$ 625k images) being the highest rated. The LAD 6.5+ is widely used by computer vision researchers to fine-tune generative image models \cite{guo2024smooth, castells2024ld}, evaluate new training methods \cite{luo2023latent, pre_train_laion_a}, and create new datasets \cite{tudosiu2024mulan, li2024laion_sg}. Additionally, Ding et al. measure the effectiveness of data poisoning attacks on text-to-image models by defining ``generation aesthetics'' as the ``\% of images with [LAP] aesthetics > 6.5'' \cite{ding2024understanding_data_poison}. In Tables \ref{tab:met_department} and \ref{tab:wikiart_dataset} we also show the number of images rated 6+ to demonstrate that the aesthetic biases we identify across the 6.5+ threshold can also be seen across the 6+ threshold.

\subsection{Findings: The LAION-Aesthetics Dataset}

LAD consists of all images from LAION-5B with English captions that LAP rated at least 4.5. In LAD, images are not stored directly. Rather, the dataset contains URLs to images hosted on websites across the internet. Most relevant for our findings, the metadata for each image in LAD contains its LAP predicted aesthetic score and a caption derived from the scraped image's alternative text. We find that LAP rates images from independent visual artists and photographers highly. Also, images with captions mentioning women, Hindus, and Christians are more likely to be included in LAP 6.5+, while images with captions mentioning men, Jews, or Muslims are more likely to be filtered out.

\begin{figure}
    \centering
    \includegraphics[width=0.8\textwidth]{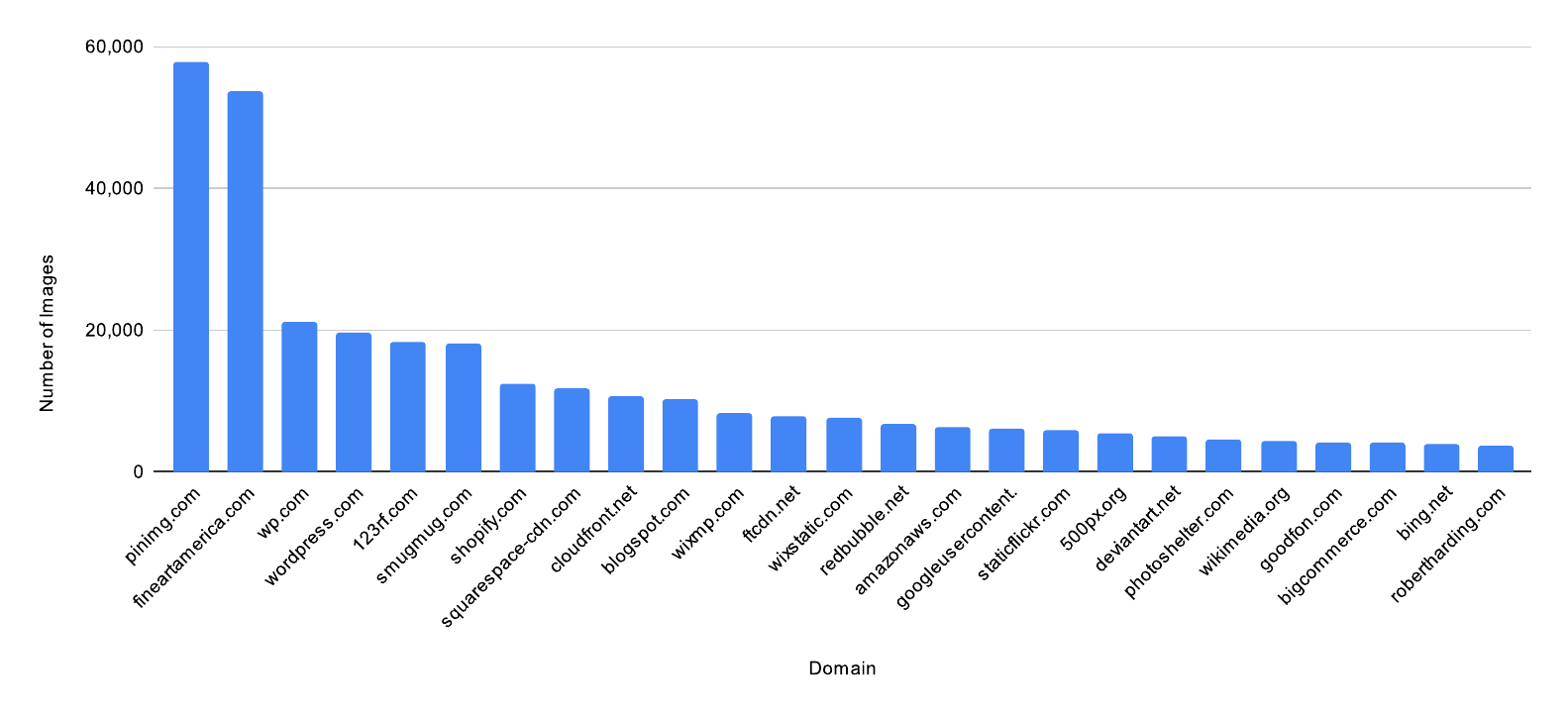}
    \caption{Top 25 domains of images in the LAION-Aesthetics Dataset rated 6.5+ by the LAION-Aesthetics Prediction model.}
    \label{fig:laion_65_url}
\end{figure}

\begin{figure}
    \centering
    \includegraphics[width=0.8\textwidth]{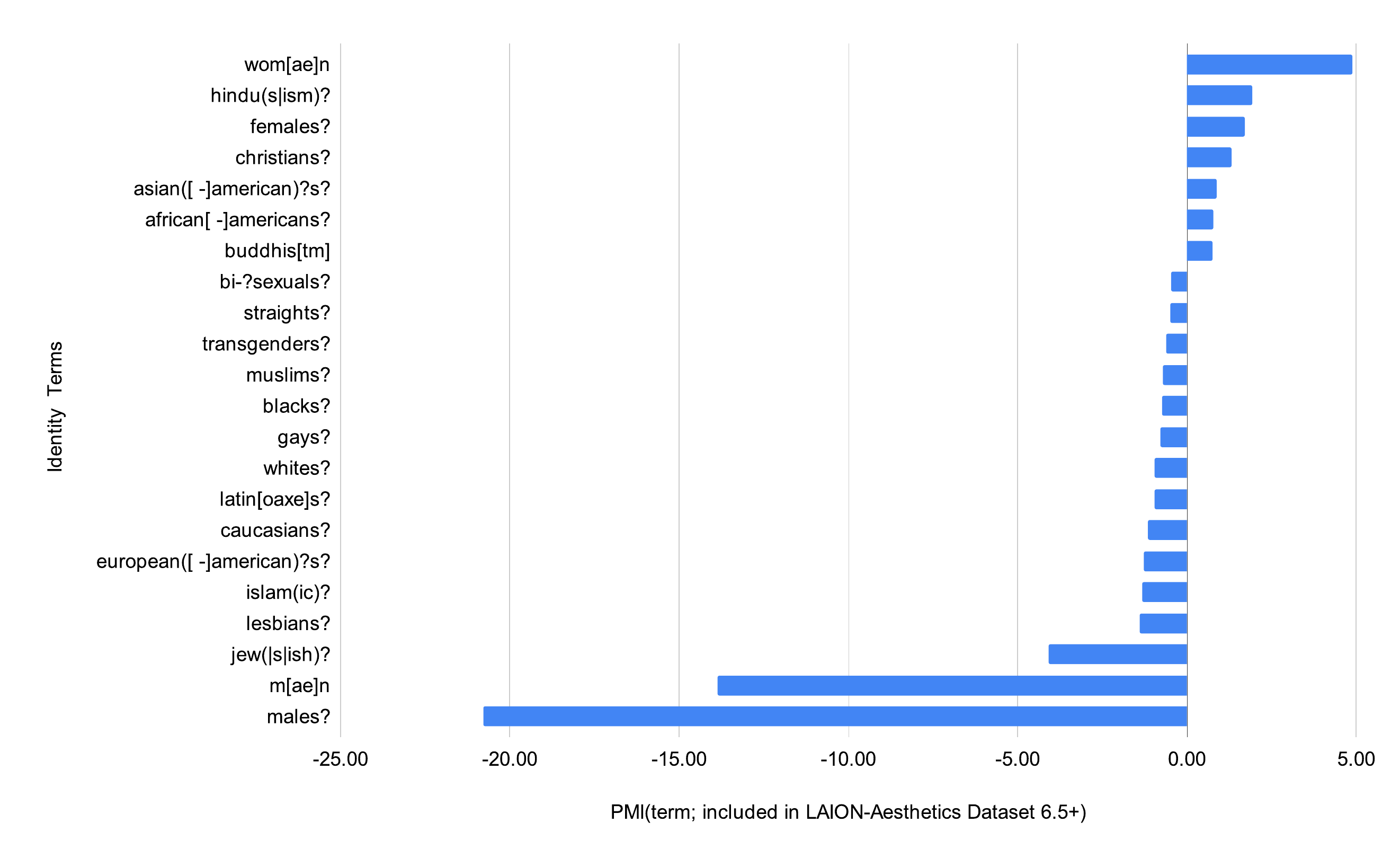}
    \caption{Pointwise Mutual Information (PMI) between regex and images being included in the 6.5+ subset of the LAION-Aesthetics Dataset. Terms with higher PMI (e.g., wom[ae]n) have higher likelihood of being included. The exact regex for each identity term was of the form \textbackslash b(?i)gays\textbackslash b.}
    \label{fig:pmi_chart}
\end{figure} 

\subsubsection{The LAION-Aesthetics Predictor highly rates images from websites used by independent visual artists and photographers}

We find substantial differences between the URLs of images below and above the 6.5 aesthetic threshold. In LAD 4.5+, the top 25 most common domains account for about 36\% of all the images in the dataset. These domains are primarily from e-commerce websites (e.g., Shopify), stock image sites (e.g., Getty Images), and web hosts (e.g., Wordpress). For LAD 6.5+, the top 25 image domains account for proximately 52\% of the dataset, meaning a majority of the images in this dataset come from the handful of websites in Figure \ref{fig:laion_65_url}. Of these 25 domains, 13 were also in the top 25 domains for LAD 4.5+. However, the other domains are primarily from smaller websites where independent visual artists (e.g., Redbubble, DeviantArt) and photographers (e.g., SmugMug, Flikr, 500px) share or sell their work. Additionally, the 25th most common source of images in LAD 6.5+ are from a small nature photography agency, robertharding.com. This indicates that LAP may rate photographic images more highly.

\begin{table*}[h]
    \centering
    
    \begin{tabular}{|p{6cm}|p{2cm}|p{2cm}|p{2cm}|p{2cm}|}
    
        \hline
        \textbf{Department Name} & \textbf{Median \newline Score} & \textbf{Num Images \newline Scored $\geq$ 6.5} & \textbf{Num Images \newline Scored $\geq$ 6} & \textbf{Num Images} \\
        \hline
        \textbf{European Paintings} & \textbf{5.73} & \textbf{35} & \textbf{483} & \textbf{2,289} \\
        \hline

        \hline
        
        \textbf{Photographs} & \textbf{5.40} & \textbf{15} & \textbf{548} & \textbf{6,685} \\
        \hline

        \hline
        \textbf{Drawings \& Prints} & \textbf{5.06} & \textbf{13} & \textbf{435} & \textbf{60,889} \\
        \hline

        \hline
        Robert Lehman Collection & 5.04 & 3 & 41 & 2,273 \\

        \hline
        The Cloisters & 4.96 & 0 & 1 & 2,258 \\
        \hline

        \hline
        Costume Institute & 4.94 & 0 & 7 & 8,124 \\
        \hline

        \hline
        \textbf{Asian Art} & \textbf{4.93} & \textbf{60} & \textbf{1,240} & \textbf{30,597} \\
        \hline

        \hline
        \textbf{The American Wing} & \textbf{4.86} & \textbf{49} & \textbf{474} & \textbf{12,292} \\
        \hline

        \hline
        Arms \& Armor & 4.82 & 0 & 2 & 7,260 \\
        \hline

        \hline
        The Libraries & 4.77 & 0 & 0 & 107 \\
        \hline

        \hline
        Musical Instruments & 4.75 & 0 & 3 & 2,498 \\
        \hline

        \hline
        Modern \& Contemporary Art & 4.73 & 1 & 11 & 865 \\
        \hline

        \hline
        European Sculpture \& Decorative Arts & 4.72 & 1 & 12 & 37,709 \\
        \hline

        \hline
        Arts of Africa, the Ancient Americas, \& Oceania & 4.71 & 0 & 0 & 6,191 \\
        \hline

        \hline
        Medieval Art & 4.70 & 0 & 1 & 6,887 \\
        \hline

        \hline
        Egyptian Art & 4.65 & 0 & 3 & 13,269 \\
        \hline

        \hline
        Islamic Art & 4.56 & 0 & 15 & 13,054 \\
        \hline

        \hline
        Ancient West Asian Art & 4.41 & 0 & 1 & 6,093 \\

        \hline

        \hline
        Greek \& Roman Art & 4.29 & 0 & 7 & 30,011 \\
        \hline

    \end{tabular}
    \caption{Predicted aesthetic score of images from the Metropolitan Museum of Art by department.}
    \label{tab:met_department}
\end{table*}

\subsubsection{The LAION-Aesthetics Predictor is more likely to rate images with descriptions mentioning woman $\geq$ 6.5, while those mentioning men or LGBTQ+ people are more likely to be filtered out.}

To understand the content of images in LAD, we compared the frequencies of words in image captions above and below the 6.5 aesthetics score threshold. Similar to Dodge et al.'s prior audits of generative AI dataset filtering \cite{dodge2021documenting}, we computed the pointwise mutual information (PMI) \cite{church1990word} of a regular expression (regex) appearing in the 6.5+ subset of images in LAD. At a high level, PMI in this case measures the probability of observing a search term in the description for an image rated 6.5+ relative to chance. Therefore, a PMI much greater than 0 suggests the search term appears more often than by chance and a PMI much less than 0 suggests the regex occurs less often than expected. Building on identity-related regexes used in prior audits \cite{dodge2021documenting, hong2024datacomp}, we computed the PMI of the regexes in Figure \ref{fig:pmi_chart} appearing in LAD 6.5+ image descriptions.

We find that images with descriptions mentioning women are more likely to be included in LAD 6.5+, while images mentioning men or LGBTQ+ communities are more likely to be excluded. Also, images with descriptions mentioning Hindu, Buddhist, or Christian communities are more likely to be included in LAD 6.5+, while images with descriptions mentioning Jews or Muslims are more likely to be filtered out. Images mentioning Asian or African communities are more likely to be included in LAD 6.5+ than those mentioning Latinx/e, Caucasian, or European communities. This may suggest that LAP more highly rates images of Asian and African communities than European communities. Alternatively, images containing Caucasians or Europeans may not mention this detail in their captions because whiteness is an unmarked racial category \cite{brekhus1998sunmarked_race}.

As PMI is a relative measure, one must also consider the \textit{absolute} number of image descriptions per regex. For example, the PMI of `whites?' and `latin[oaxe]s?' are both -0.95. However, there are 9,617 images in LAD 6.5+ with `whites?' in their description but only 26 images for ther regex `latin[oaxe]s?'. Not only does `wom[ae]n' have the highest PMI but the regex also appears in more image descriptions in LAD 6.5+ than any other regex in Figure \ref{fig:pmi_chart} (15,706). Meanwhile, regexes associated with LGBTQ+ identities are far rarer in LAD 6.5+ image descriptions: `gays?' (102), `lesbians?' (20), `transgenders?' (9), `bi-?sexuals?' (2), and `non[ -]?binary' (0). Understanding the cultural implications of aesthetic evaluation requires considering more than the representation of identity groups \cite{taylor2025straightening}. Below, we explore aesthetic filtering in greater detail by examining how LAP rates art from various cultures.

\begin{table*}[h]
    \centering
    
    \begin{tabular}{|p{2cm}|p{3.5cm}|p{2cm}|p{2cm}|p{2cm}|p{2cm}|}
    
        \hline
        &  & \textbf{Median Score} & \textbf{Num Images \newline Scored $\geq$ 6.5} & \textbf{Num Images \newline Scored $\geq$ 6} & \textbf{Num Images} \\

        \hline
        
        \textbf{Genre} & Cityscape & 5.97 & 539 & 2,151 & 4,602 \\
        \hline
        & Portrait & 5.90 & 693 & 5,616 & 14,112 \\
        \hline
        & Landscape & 5.87 & 505 & 4,637 & 13,358 \\
        \hline
        & ... & ... & ... & ... & ... \\

        \hline
        & Illustration & 5.36 & 11 & 140 & 1,902 \\
        \hline
        & Nude Painting & 5.45 & 3 & 117 & 1,923 \\
        \hline
        & Abstract Painting & 4.94 & 0 & 0 & 4,968 \\

        \hline
        \hline

        \textbf{Style} & Realism & 5.87 & 634 & 4,155 & 10,733 \\
        \hline
        & Romanticism & 5.87 & 371 & 2,723 & 7,019 \\
        \hline
        & Impressionism & 5.89 & 646 & 4,898 & 13,060 \\
        \hline
        & ... & ... & ... & ... & ... \\
        \hline
        & Cubism & 5.50 & 12 & 210 & 2,235 \\
        \hline
        & Pop Art & 5.05 & 2 & 33 & 1,483 \\
        \hline
        & Abstract Expressionism & 4.96 & 0 & 7 & 2782 \\

        \hline
        \hline

        \textbf{Artist} & Antoine Blanchard & 6.94 & 168 & 170 & 170 \\
        \hline
        & Edouard Cortes & 7.09 & 208 & 214 & 214 \\
        \hline

        & ... & ... & ... & ... & ... \\
        
        \hline
        & Pablo Picasso & 5.51 & 1 & 65 & 762 \\
        \hline
        & Andy Warhol & 5.22 & 0 & 6 & 107 \\
        \hline
        & Salvador Dalí & 5.17 & 0 & 20 & 479 \\

        \hline

    \end{tabular}
    \caption{A representative selection of the highest and lowest scored genres, styles, and artists in WikiArt. The full results per genre, style, and artist can be downloaded at  https://github.com/jtaylor351/laion\_aesthetics\_audit\_data/tree/main}
    \label{tab:wikiart_dataset}
\end{table*}

\subsection{Findings: The Metropolitan Museum of Art Dataset}

To examine how the LAP evaluates art across different mediums and cultures, we scored the 249,351 images from the MET dataset using LAP. We find LAP rates two-dimensional art by western or Japanese artists most highly.

\subsubsection{The LAION-Aesthetics Predictor rates western art more highly than art from the Majority World}
 As seen in Table \ref{tab:met_department}, the departments with images scored $\geq$ 6.5 are primarily European Paintings, Photographs and Drawing \& Prints as well as art from Europe, Asia, and America. On the other hand, not a single piece of African, Native American, Oceanian, Egyptian, Islamic, Ancient West Asian, and Greek \& Roman Art was rated 6.5+. This bias is even more apparent when looking at the images rated 6+. \textbf{Approximately 97\% of all images scored 6+ from the MET (3180 of 3284) come from only five departments: (1) Asian Art, (2) Photographs, (3) European Paintings, (4) The American Wing, and (5) Drawing and Prints.} This may be partially because the LAP privileges figurative, two-dimensional art, which would also explain why more abstract art (Modern \& Contemporary Art) and physical artifacts (European Sculpture \& Decorative Arts) scored lower.

\subsubsection{The LAION-Aesthetics Predictor rates paintings and photographs by western or Japanese artists most highly.} Across the 249,351 images corresponding to 37,878 unique mediums and 14,946 known artists in the MET dataset, only 177 images across 35 mediums and 99 known artists were rated as $\geq$ 6.5 by LAP. After manually combining similar mediums (e.g., `watercolor' and `watercolor on paper' ), we found that the MET images rated $\geq$ 6.5 come from only 5 mediums: 72 oil paintings, 60 Japanese woodblock prints, 26 watercolor paintings, 15 photographs, 2 drawings, 1 piece of stained glass, and 1 tempera painting (a predecessor to oil paint). A total of 14 artists had more than one piece rated $\geq$ 6.5, accounting for 88 (50\%) of the 177 top rated images.

A manual inspection indicates that images rated $\geq$ 6.5 are typically two-dimensional realistic representations of landscapes (e.g., paintings of mountains) or portraits (e.g., photographs or paintings of people). This can also be seen in the artists with multiple images in the top 177. The top two artists with the most images rated $\geq$ 6.5, Katsushika Hokusai (32) and Utagawa Hiroshige (23), are mid-19th century ukiyo-e woodblock print artists who focus on Japanese landscapes \cite{kafu2012ukiyo}. Likewise, the artists with the third and sixth most images rated $\geq$ 6.5 are the landscape painters William Trost Richards (6) and Albert Bierstadt (3). The artists with the fourth (Southworth and Hawes, 4) and fifth (Suzuki Shin'ichi, 4) most images in the top 177 are portrait photographers. While the MET dataset contains a broad range of art across cultures and mediums, those that LAP rated $\geq$ 6.5 were mostly paintings, prints, or photographs by western or Japanese artists. In the next section, we examine how LAP evaluates paintings in greater detail through the WikiArt dataset.

\subsection{Findings: The WikiArt Dataset}

The WikiArt dataset consists of 81,444 pieces of visual art from 128 known artists (excluding unknown artists), primarily western artists from the mid-1800s to mid-1900s. Each piece is assigned one of 27 styles across 10 genres. Genres describe the content of a piece (e.g., landscape or portrait) meanwhile style describes the piece's form (e.g., realism or impressionism).

\subsubsection{The LAION-Aesthetics Predictor more highly rates realistic images of cityscapes, landscapes, and portraits, while less realistic art was rated lower.} As seen in Table \ref{tab:wikiart_dataset}, we find that LAP highly rates visual art of Cityscapes, Landscapes, and Portraits. These three genres account for 39\% of all images in WikiArt, but these genres make up 73\% of all images rated 6.5+. Likewise, we found that more realistic and representational styles, such as Realism, Romanticism, and Impressionism were highly rated. Here, both style and genre are linked as portraits, landscapes, and cityscape are common subjects in realist, romantic, and impressionistic art. On the other hand, LAP rates less representational art, such as cubism or abstract art, lower. The consequences of this aesthetic bias can be seen across artists. Nearly every piece by the french cityscape artists Edouard Cortès (208 of 214) and Antoine Blanchard (168 of 170) were rated 6.5+. In contrast, more influential but less realistic artists—Pablo Picasso, Andy Warhol, and Salvador Dalí—were rated lower.

\begin{table*}[h]
    \centering
    
    \begin{tabular}{|>{\centering}m{2.5cm}|>{\centering}m{3.5cm}|>{\centering}m{3.5cm}|>{\centering\arraybackslash}m{3.5cm}|}
    
        \hline
         \textbf{Aesthetics Score \newline Range} & \textbf{LAION-Aesthetics \newline Dataset} & \textbf{Met Dataset} & \textbf{WikiArt Dataset} \\
        \hline

        \textbf{6.5 $\leq$ score}   & \includegraphics[height=1.5cm]{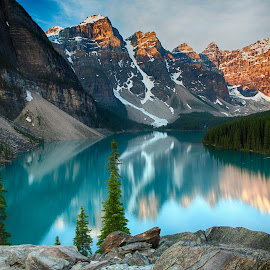}    \Description[A Mountain]{A photo of a mountain and a lake} & \includegraphics[height=1.5cm]{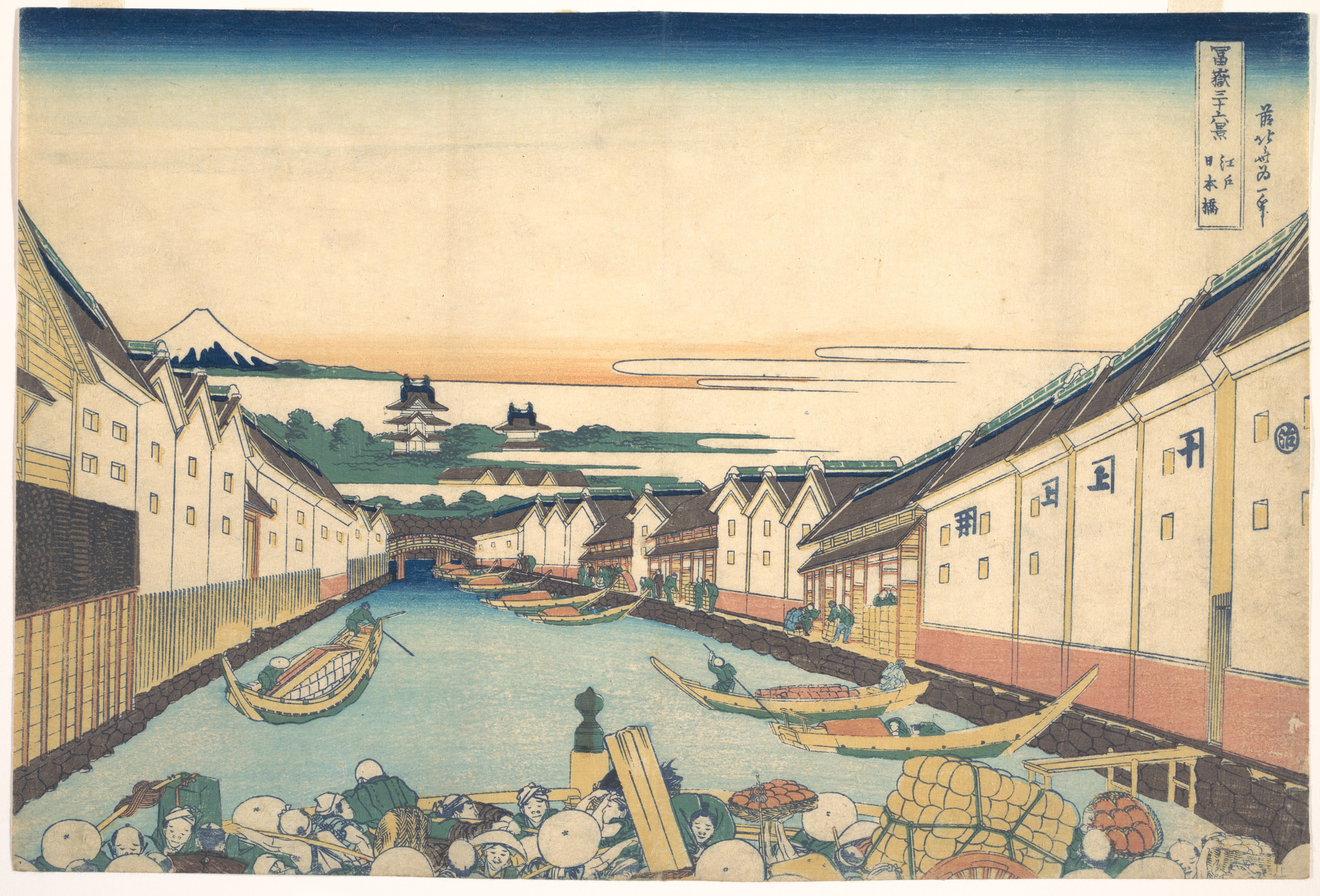} \Description[A Japanese Landscape]{A Japanese woodblock print of a river with boats and a mountain in the background} & \includegraphics[height=1.5cm]{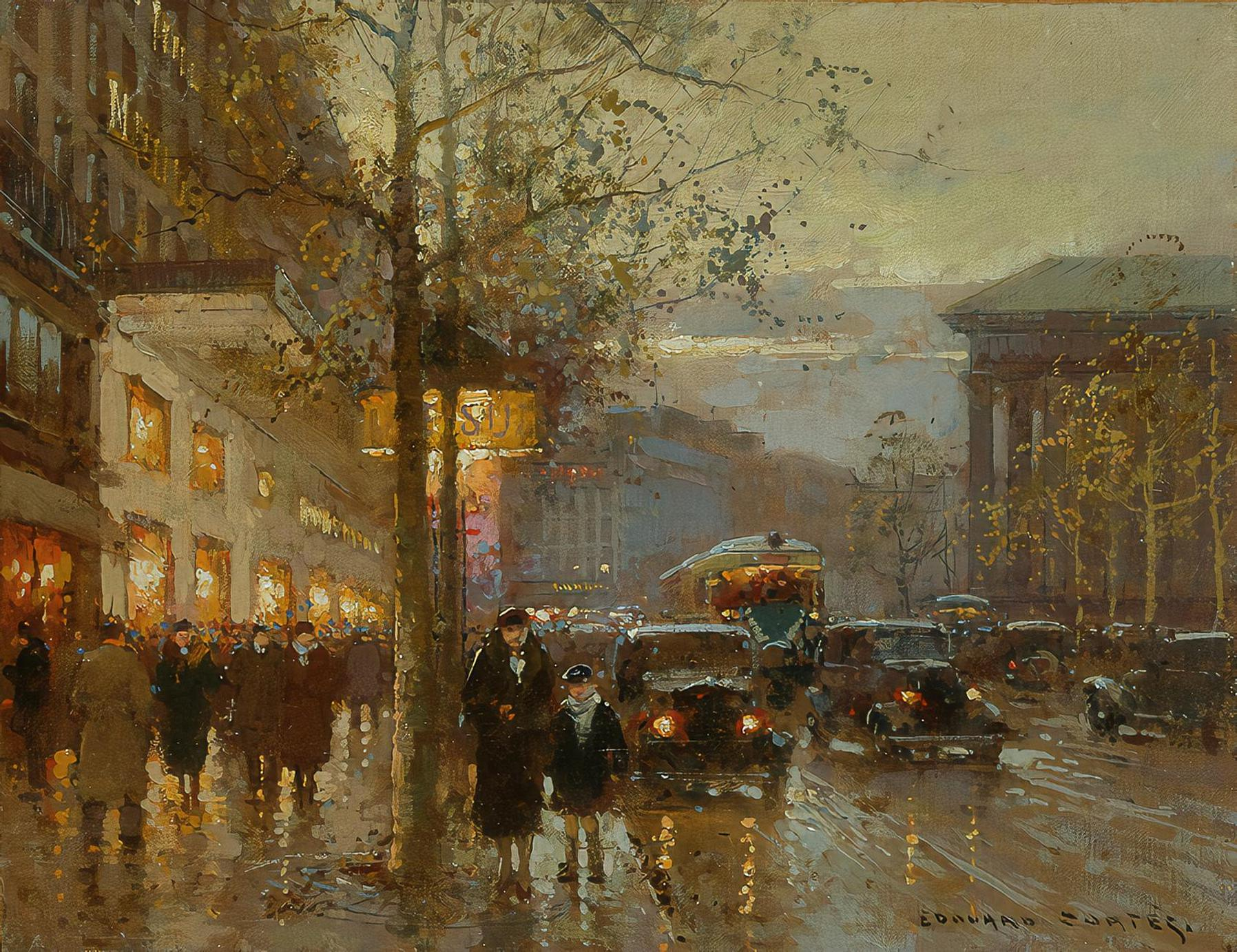} \Description[A cityscape painting]{A post-impressionist painting of a cityscape with cars and people walking} \\ \hline 

        \textbf{5.5 $\leq$ score $<$ 6.5}   & \includegraphics[height=1.5cm]{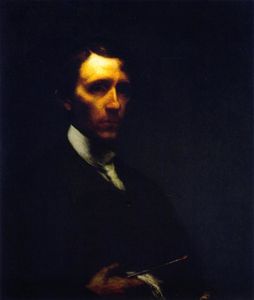} \Description{Painting of a white man} & \includegraphics[height=1.5cm]{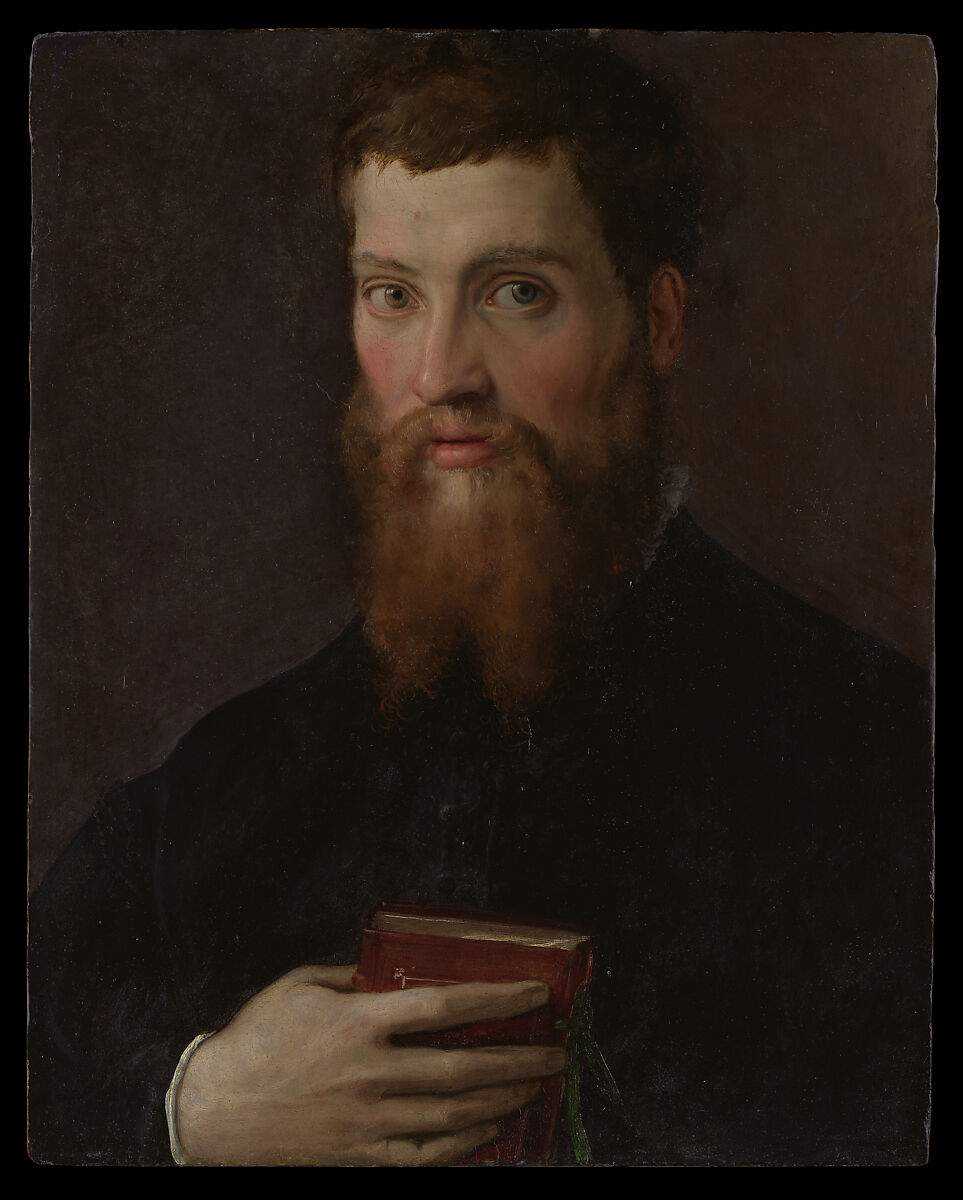} \Description{Painting of a white man} & \includegraphics[height=1.5cm]{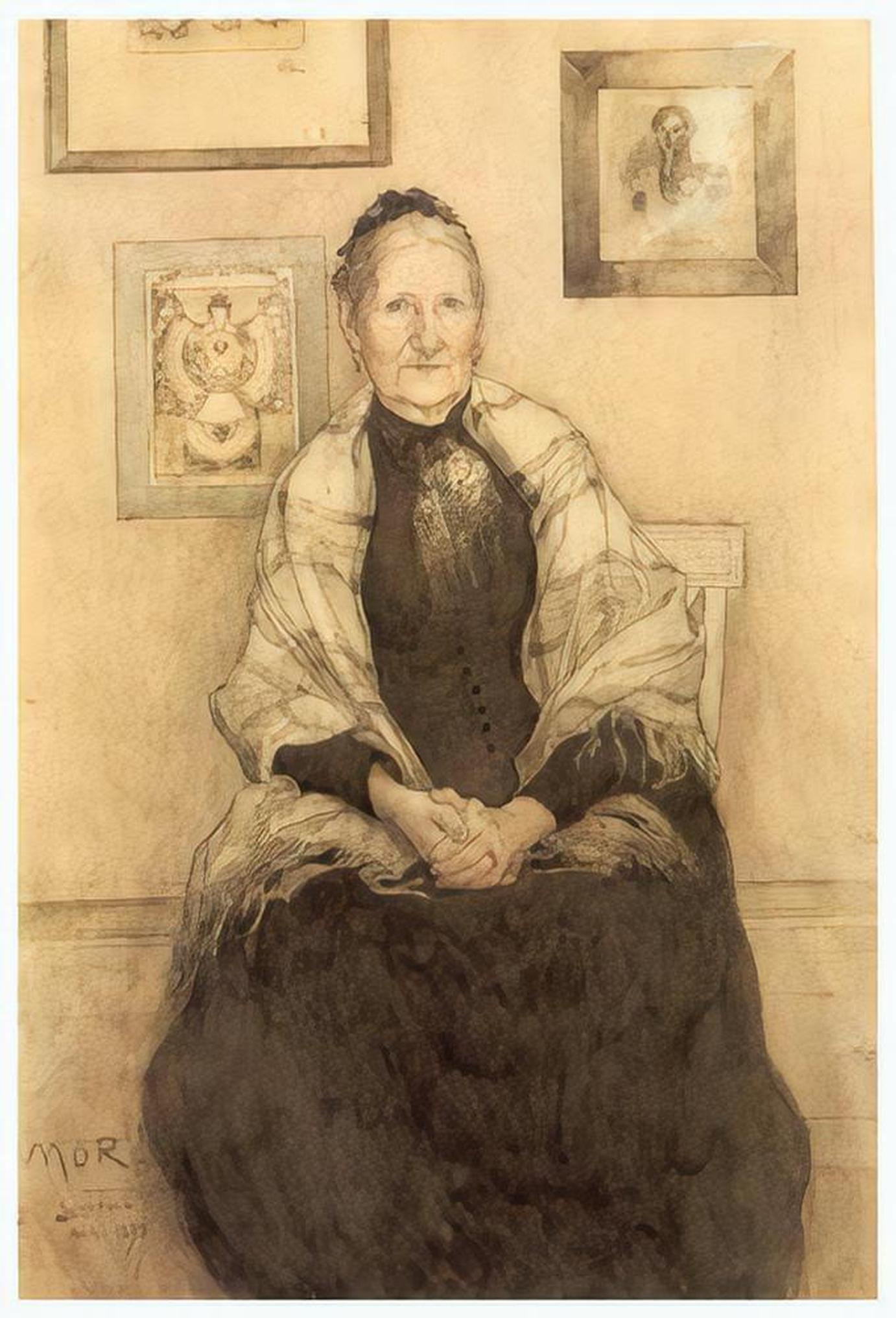} \Description{Painting of an older white woman} \\ \hline

        \textbf{4.5 $\leq$ score $<$ 5.5}   & \includegraphics[height=1.5cm]{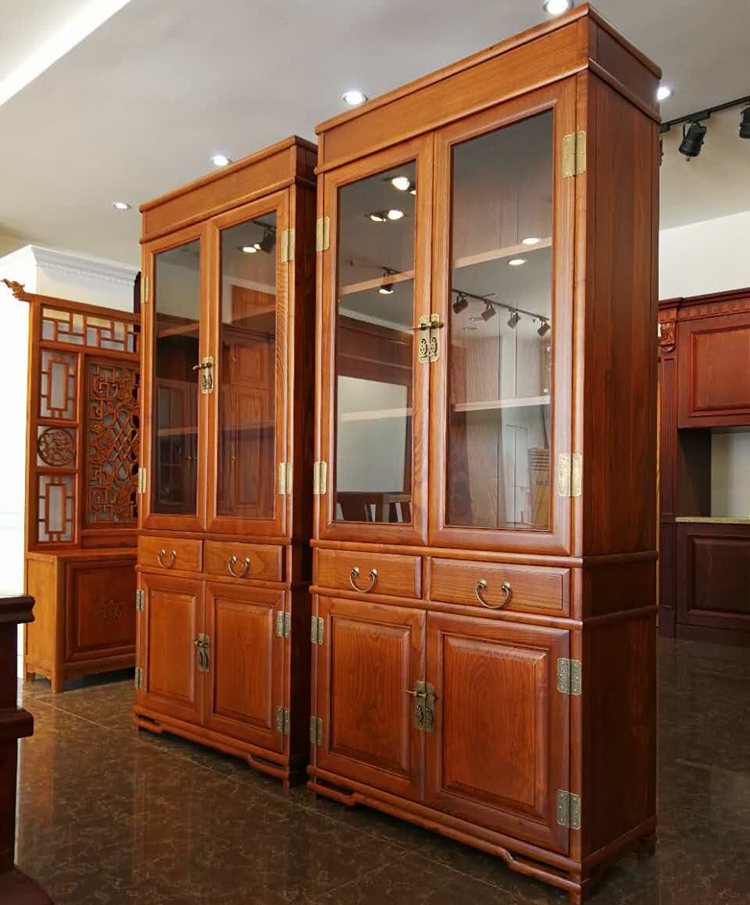} \Description{A wooden wall unit with glass doors} & \includegraphics[height=1.5cm]{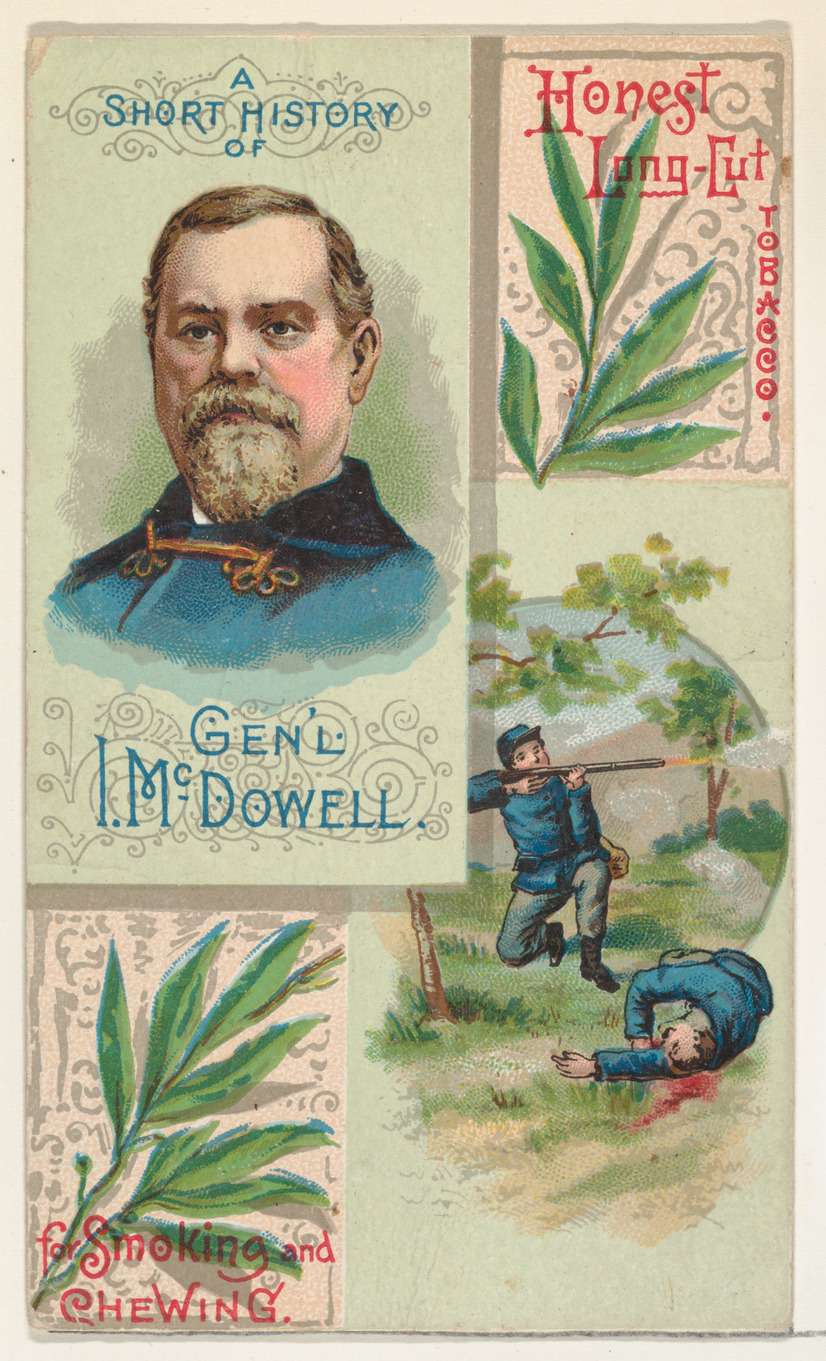} \Description[Tobacco Trading Card]{An illustrated trading card to promote Honest Long Cut Smoking and Chewing Tobacco} & \includegraphics[height=1.5cm]{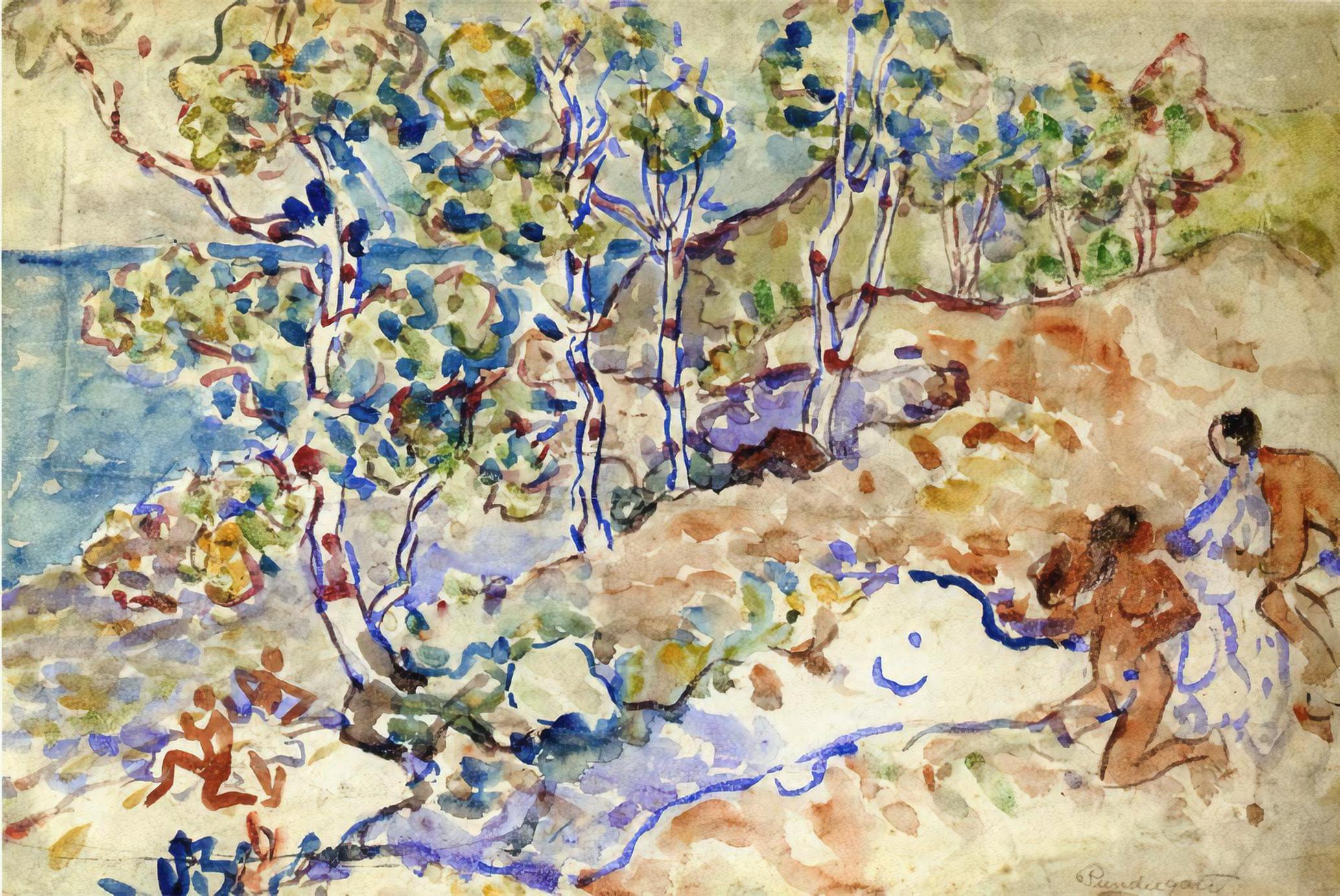} \Description[A watercolor painting]{A post-impresionistic watercolor painting of people on a beach amid trees } \\ \hline

        \textbf{score $<$ 4.5}  &  Not Applicable & \includegraphics[height=1.5cm]{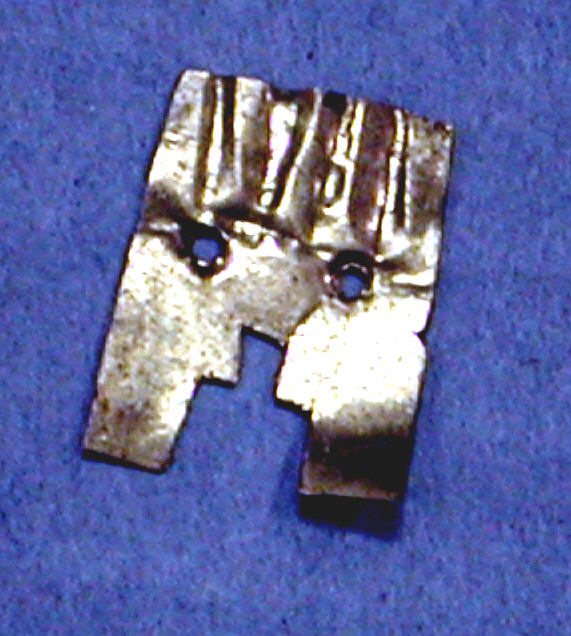} \Description{A hammered gold artwork} & \includegraphics[height=1.5cm]{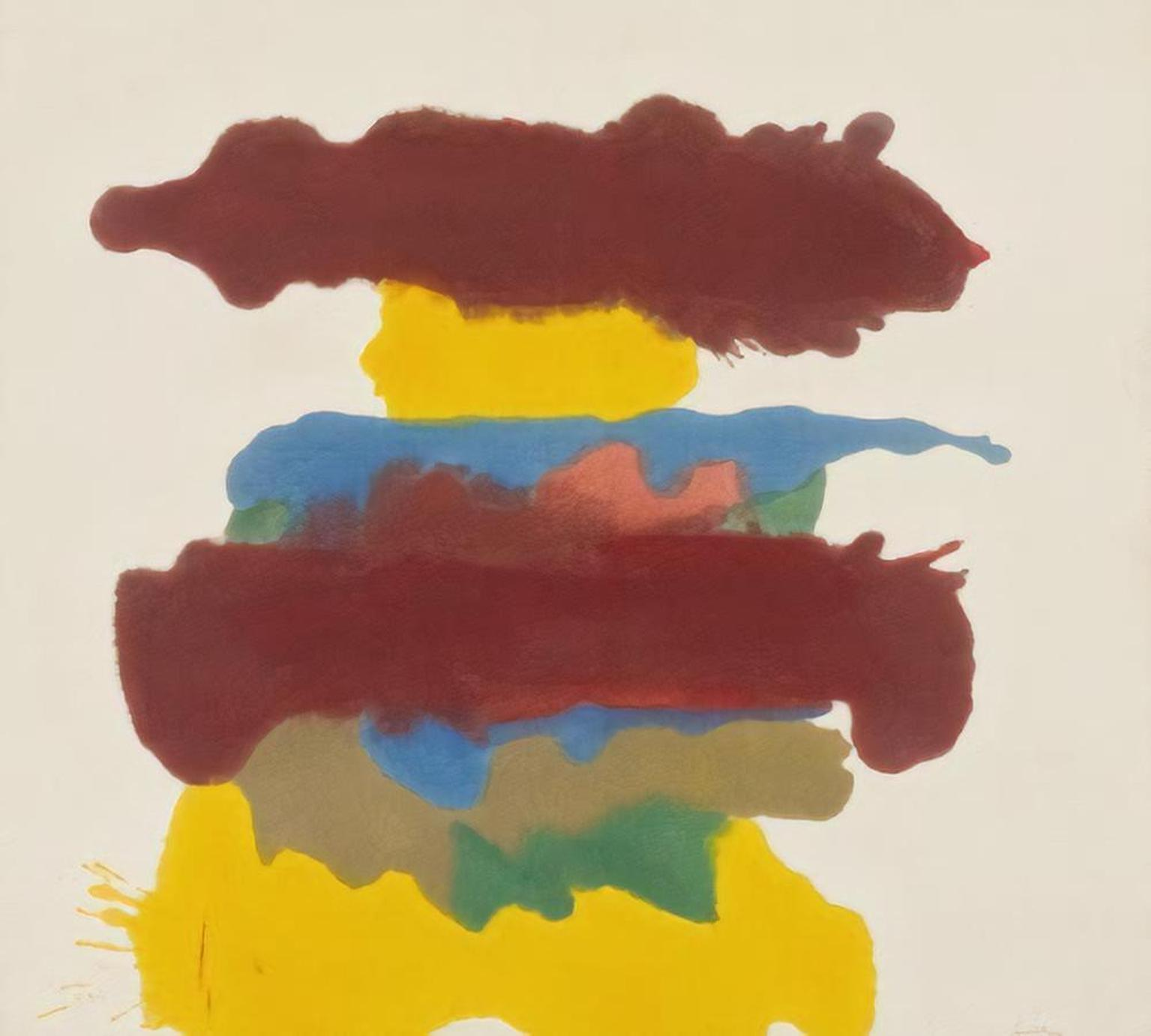} \Description{An abstract painting with a white background and red, yellow, and blue horizontal marks.} \\ \hline 
        
    \end{tabular}
    \caption{A selection of images across the three datasets showing that realistic images of landscapes and cities are most highly rated.}
    \label{tab:aesthetic_score_example_images}
\end{table*}

\subsection{Findings: Qualitative Visual Analysis}

To augment our quantitative analysis, we randomly sampled 5 images from each dataset in the ranges of: [7,), [6.5, 7.0) ... [4.5, 5.0), (, 4.5). Given no images in LAD are rated $<$ 4.5 and two images in the MET were rated $\geq$ 7, this yielded 30 images from LAD, 32 images from the MET, and 35 images from WikiArt. A representative image from each dataset across different score ranges can be seen in Table \ref{tab:aesthetic_score_example_images}. \textbf{Our inspection of these images further suggests that LAP tends to rates realistic images of cityscapes, landscapes, and people higher than images of three-dimensional objects or abstract art.} Among the images rated 6.5+ across the datasets, all are of landscapes/nature (11/27), cityscapes (9/27), or people (7/27). Between 5.5 and 6.5, most images are of people (20/30) or landscapes/nature (5/30). The images scored below 5.5 are primarily of inanimate objects (20/40), people (10/40), or abstract art (7/40).

\section{How the LAION-Aesthetics Predictor was Made}
\label{sec:ethnography}

\citet{seaver2017algorithms} suggests that audits can demonstrate the \textit{existence} of biases, but ``what [audits] cannot do is explain conclusively how that disparate impact came about.'' In this section, we explore the potential origins of the aesthetic biases described above by conducting a trace ethnography of how the LAP was made.

\subsection{Trace Ethnographic Methodology}

Ethnography is a qualitative method from anthropology seeking to explain how groups of people interpret and act in the world \cite{dourish2014reading_ethnography}. Ethnographic research typically involves extended participant observation and interviews within physical field-sites, ranging from remote villages \cite{geertz1973interpretation} to AI research labs \cite{forsythe2001studying}. While ethnography began within physical field sites, the 1990s saw a proliferation of multi-sited ethnographic research due to the rise of globalization and the internet \cite{marcus1995ethnography}. Within this paradigm, \citet{geiger2011trace} propose trace ethnography as ``a powerful and flexible methodology, able to turn thin documentary traces into ‘thick descriptions’ \cite{geertz1973interpretation} of actors and events that are often invisible in today's distributed, networked environments.'' For example, \citet{geiger2011trace} reconstruct the blocking of a Wikipedia vandal by drawing on thin documentary traces of edit history. 

We conducted a trace ethnography of the development of the LAION-Aesthetics Predictor by leveraging a variety of digital documents. Specifically, we draw on: (1) the LAION-Aesthetics announcement blog post \cite{laion_aesthetics}, (2) the code and data repositories hosting the LAP model,\footnote{https://github.com/christophschuhmann/improved-aesthetic-predictor} SAC dataset,\footnote{https://github.com/JD-P/simulacra-aesthetic-captions} AVA dataset,\footnote{https://github.com/imfing/ava\_downloader} and LAION-Logos dataset,\footnote{https://huggingface.co/datasets/ChristophSchuhmann/aesthetic-logo-ratings} (3) the academic paper for the AVA dataset \cite{murray2012ava}, and (4) a $\sim$10 minute YouTube video explaining how LAP was made.\footnote{https://www.youtube.com/watch?v=OZ3krwyoeWw} Our analysis (summarized in Table \ref{tab:aesthetic_datasets_grid}) focused on understanding how LAP and its training data were created and where the images and annotations in LAP's training data come from.

\subsection{Findings: How the LAION-Aesthetics Predictor was Created}

    \subsubsection{Model development choices were made according to the individual taste of LAION's founder}
    
    According to the aforementioned YouTube video, the founder of LAION (Christoph Schuhmann) created LAP on his own. LAP is a multi-layer perceptron model with linear layers trained to predict the aesthetic score of an image based on a CLIP embedding.\footnote{CLIP models are commonly used in computer vision to semantically represent both texts and images in the same vector space \cite{radford2021clip}}  Although he experimented with other models, Schuhmann chose this architecture because \textit{``for [his] use case the simple linear layer worked best just from visual inspection.''} Schuhmann trained LAP using three distinct datasets, each of which measures the aesthetic quality of an image on a scale from 1 to 10: (1) Aesthetic Visual Analysis (AVA), (2) Simulacra Aesthetic Captions (SAC), and (3) LAION-Logos. Once again, Schuhmann explains that he chose to weight each dataset equally in the training process because \textit{``from [his] subjective taste it was best to just have every dataset once in the training data.''} In other words, Schuhmann decided how to train LAP based largely on his individual taste. 
    
    Schuhmann seems to recognize various limitations of LAP. He encourages others to \textit{``experiment''} with different model architectures, newer CLIP embedding models, and larger aesthetic evaluation datasets. Additionally, Schuhmann asks for volunteers to reproduce the training data curation steps described in the video because \textit{``[he] had some ugly code somewhere but [he] can’t find it anymore.''} He also asks for volunteers in the last comment of the explainer video: \textit{``This is all an ongoing area of research and it would be really cool to have someone working on this.''} Schuhmann appears to recognize the slapdash way LAP was made and expects that it will be improved upon in the future. However, at the time of our writing in January 2026, LAION has not released a new aesthetic evaluation model. 

    \begin{table*}[h]
    \centering
    
    \begin{tabular}{|p{4cm}|p{3.5cm}|p{3cm}|p{3cm}|}
    
        \hline
         & \textbf{AVA} & \textbf{SAC} & \textbf{LAION-Logos} \\
        \hline
        \textbf{Year Released} & 2012 & 2022 & 2022 \\
        \hline
        \textbf{Number of Images} & 255,530 & 146,372 & 26,730 \\
        \hline
        \textbf{Mean Ratings per Image} & 210.1 & 1.2 & 1.48 \\
        \hline
        \textbf{Median Ratings per Image} & 201 & 1 & 1 \\
        \hline
        \textbf{Images data consent} & No & Yes & No \\
        \hline
        \textbf{Ratings data consent} & No & Yes & Yes \\
        % \hline
        \hline
        \textbf{Dataset Documentation} & Academic paper & GitHub README.md & No documentation \\
        \hline
        \textbf{Where \textit{images} come from} & www.dpchallenge.com & T2I generated & LAION-5B \\
        \hline
        \textbf{Where \textit{scores} come from} & www.dpchallenge.com & 294 AI enthusiasts & 18 AI enthusiasts \\
        \hline
        \textbf{What scores \textit{measure}} & \textbf{Relative} rating on a theme & \textbf{Absolute} rating & \textbf{Absolute} rating \\
        \hline
    \end{tabular}
    \caption{Comparison of Datasets Used to Train the LAION-Aesthetics Predictor}
    \label{tab:aesthetic_datasets_grid}
\end{table*}

    \subsubsection{Training datasets are inconsistently documented}

    AVA is an influential predecessor to the LAION-Aesthetics dataset within the computer vision community \cite{ramesh2022hierarchical, wang2023exploring, khosla2015understanding, lu2014rapid}. The computer vision researchers Murray et al. collected AVA by scraping images and ratings from an online photography competition website (www.dpchallenge.com) in 2012 \cite{murray2012ava}. SAC was released in July of 2022 by John David Pressman, a Stability AI researcher at the time \cite{pressmancrowson2022}. Pressman collected this dataset using a Discord bot shared in AI-art communities that allowed volunteers to generate images and vote on their perceived aesthetic quality. Although we were able to find the content of each dataset, the quality of their documentation varies widely. AVA was the most well documented because it was published at a computer vision conference \cite{murray2012ava}. While SAC was not formally published in an academic setting, its GitHub repository has an extensive README.md file that functions as a data card \cite{pushkarna2022data_card}. As we will describe next, LAION-Logos had little documentation.

    We initially struggled to find LAION-Logos. While AVA and SAC are linked in the LAION-Aesthetics blog post, LAION-Logos is not. Searching ``LAION-Logos'' on Google yielded only websites and academic articles alluding to the original blog. In response, we searched the public LAION Discord, eventually finding an unlisted YouTube video posted by Schuhmann in 2022.\footnote{https://www.youtube.com/watch?v=OZ3krwyoeWw} Because this video is unlisted, it is not searchable on YouTube or indexed by search engines. In the video, Schuhmann explains that LAION-Logos is located on Hugging Face.\footnote{https://huggingface.co/datasets/ChristophSchuhmann/aesthetic-logo-ratings} LAION-Logos is sparsely documented with a one sentence explanation and rarely used, downloaded only 40 times in the month prior at the time of our writing. While LAION-Logos is technically publicly available, our challenges finding the dateset suggests that it has been \textit{de facto} private. To the best of our knowledge, we are the first to document LAION-Logos.

    \subsubsection{Training datasets were created for different purposes across different time periods}

    LAION-Logos and SAC were created to improve generative image models. LAION-Logos was collected to \textit{``improve the [LAION aesthetic prediction] models abilities to evaluate images with more or less aesthetic texts in them''} \cite{laion_aesthetics} While the SAC documentation says the dateset can be used in multiple ways, like ``Filtering Datasets'' and ``Alignment Research,'' its documentation indicates that SAC was primarily collected for LAION-Aesthetics. This is evident in the acknowledgment: \textit{``Special thanks to [name] for his early feedback and suggestions on the bot, as well as writing the initial public domain prompt and rating dataset that allowed us to make LAION-aesthetic as early as we did''}
    
    In contrast, AVA was collected a decade prior to the popularization of generative AI research. To motivate the construction of AVA, \citet{murray2012ava} suggest that \textit{``With the ever-expanding volume of visual content available, the ability to organize and navigate such content by aesthetic preference is becoming increasingly important.''} The authors position AVA as a solution to the problem of information overload, a common refrain in recommendation research \cite{seaver2022computing_taste}. Perhaps because AVA was created within a personalization paradigm, AVA embodies a more pluralistic conception of aesthetics than the universalism of LAION-Logos and SAC. We describe these competing conceptions of aesthetics below in greater detail.

\subsection{Limitations of the Images and Aesthetic Annotations Used to Train the LAION-Aesthetics Predictor}

    \subsubsection{Images and aesthetic annotations were collected with mixed levels of consent} These datasets were collected with varying levels of consent (Table \ref{tab:aesthetic_datasets_grid}). Neither the images nor the ratings in AVA were collected with the consent of dpchallenge.com members. For LAION-Logos, annotators likely knew their ratings would be used to create a dataset, but image creators did not consent. For SAC, the AI enthusiasts generating and annotating images consented to their images and annotations being used to create an aesthetic evaluation dataset.

    \subsubsection{The images used to train LAION-Aesthetics Predictor are primarily photographs.} Of the 428,632 images used to train LAP, 255,530 ($\sim$60\%) come from AVA, 146,372 ($\sim$34\%) from SAC, and 26,730 ($\sim$6\%) from LAION-Logos. Both the images and ratings from AVA are from photographers who participated in the English-language online art competition website dpchallenge.com between its founding in 2002 and its scraping in 2012. The images in SAC were generated and rated by a handful of AI enthusiasts in 2022 using an early version of Stable Diffusion, with the top 50 most active volunteers accounting for 87.4\% of all ratings. Finally, the images in LAION-Logos from LAION-5B were scraped from the internet no later than early 2022 and rated by only 18 LAION volunteers. The images used to train LAP have a number of limitations. Temporally, most images are from before 2012, meaning newer content may be underrepresented. Also, the over-representation of photographs in LAP's training data may explain why the model highly rates photorealism.

    \subsubsection{Annotators are primarily English-speaking photographers or western AI-enthusiasts.} When the AVA data was created (2002 - 2012), the World Bank estimate that less than 33\% of the world's population used the internet, most of whom were in western countries.\footnote{https://data.worldbank.org/indicator/IT.NET.USER.ZS} Therefore, the images and ratings posted on the English-language website dpchallenge.com at the time likely reflect the tastes of western photographers. The SAC README cautions: \textit{``Participants in [SAC] are largely WEIRD, or Western, Educated, Industrialized, Rich, and Democratic. This means that the aesthetic preferences recorded are not universal among humanity. While we didn't take a survey of the demographic makeup of SimulacraBot users, it can be assumed they largely reside in the United States and Europe.''} The README also warns that participants were \textit{``mostly open source AI developers and enthusiasts,'' so ``their aesthetic feedback is going to lean STEM, fantasy, nerdy, esoteric, etc.''} While the identity of LAION-Logos annotators is unclear, we expect they share similar biases to SAC. Moreover, a majority of the ratings in SAC and LAION-Logos were chosen by a single person (Table \ref{tab:aesthetic_datasets_grid}). 

    \subsubsection{The LAION-Aesthetics Predictor was trained by conflating different measurements of aesthetics across the datasets.} Per the LAION-Aesthetics blog, LAP was trained to ``predict the rating people gave when they were asked 'How much do you like this image on a scale from 1 to 10?''' \cite{laion_aesthetics}. While SAC and LAION-Logos measure aesthetics \textit{absolutely}, AVA measures aesthetics \textit{relatively}. Votes on dpchallenge do not measure the aesthetic quality of a de-contextualized image but rather the quality of an image \textit{relative} to a topic, like skyscapes. As prizes are award to images based on the average final \textit{ranking} of aesthetic votes for a challenge, votes are comparable within a topic but not between topics. \citet{murray2012ava} find that images in challenges with negative affective valences, like fear, tend to receive lower absolute scores. This implies that a model built on absolute scores rather than relative scores within challenges may be confounded by the emotional valence of an image. 

    \citet{murray2012ava} are highly critical of de-contextualized aesthetic scores: \textit{``The interpretation of these aesthetic judgments [in prior aesthetic datasets], expressed under the form of numeric scores, has always been taken for granted. Yet a deeper analysis of the context in which these judgments are given is essential. The result of this lack of context is that it is difficult to understand what the aesthetic classifiers really model when trained with such datasets.''} This is why AVA represents aesthetic scores as a distribution of votes, allowing one to examine not only the mean aesthetic vote but also the variance. The authors found that images with high variance, or greater disagreement between voters in the image competition, tended to be ``non-conventional,'' such as ``innovative application of photographic techniques.'' This nuance, however, is lost because LAP is trained by averaging the aesthetic scores.

\section{Discussion}

\subsection{The Algorithmic Gaze}

Western art has historically been made to please white, western, heterosexual, men \cite{berger2008ways, mulvey_male_gaze_1975, kaplan2012looking}. Similarly, we found that LAP was designed to the taste of one particular white, western, man: the founder of LAION. Perhaps as a result, LAP is aesthetically biased toward realistic images of western and Japanese art. Also, images with captions mentioning women in the LAION-Aesthetics Dataset were more likely to be rated 6.5+ than those mentioning men or LGBTQ+ people. Below, we discuss LAP's imperial, realist, and male algorithmic gaze.

\subsubsection{The Imperial Gaze}

In our analysis of art from the MET, all of the images rated 6.5+ were paintings, photographs and prints from western or Japanese artists. Not a single piece of African, Oceanian, Native American, Islamic, Egyptian, or West Asian art was rated 6.5+. While the highly-rated Japanese art may seem like an outlier, western artists have been influenced by a racist fascination with ``eastern'' \cite{said1979orientalism} aesthetics for centuries \cite{porter1999chinoiserie}. LAP embodies what \citet{kaplan2012looking} refers to as the imperial gaze, wherein art is judged according white, western tastes. Due to the imperial gaze of LAP, the widespread use of this model to curate training data may exacerbate existing multi-modal biases, like struggling to generate Native American \cite{taylor2025straightening} or Persian \cite{qadri2025non_western_artworld} art. 

Similar to prior work on biases in data ``quality'' filtering \cite{dodge2021documenting, hong2024datacomp}, our findings demonstrate that aesthetic evaluations may lead to representational harms by filtering out non-western art. We encourage future research into how other image evaluation methods, such as FID scores \cite{heusel2017fid} or CLIP similarity \cite{radford2021clip}, asses non-western art. At the same time, we encourage researchers to explore aesthetic biases outside of western ontologies. Notably, none of the images in the MET department of  ``Africa, the Ancient Americas, and Oceania'' scored 6+ (Table \ref{tab:met_department}). However, the consolidation of art from different continents and time periods into one department is a vestige of the colonialist category of ``primitive art'' \cite{price2001primitive}. By using the MET to evaluate LAP, our analysis inherits this coloniality. In calling for a research agenda focused on aesthetic biases, we caution against reinscribing the imperial gaze embedded in western art history.

\subsubsection{The Realist Gaze}

We found that LAP highly rates realistic and figurative images, such as landscape paintings. For centuries, one of the primary criteria for evaluating western visual art was how well it imitates reality \cite{potolsky2006mimesis}. However, the modern art diverged from this tradition by focusing on the ideas and emotions behind art \cite{fry1920essay_in_aesthetics}, often drawing inspiration from less realistic non-western art \cite{harney2018mapping}. LAP's realism bias risks devaluing both modern and non-western artistic cultures and parallels troubling historical precedents. Famously, the Degenerate Art Exhibition under Nazi Germany encouraged visitors to mock modern art \cite{degenerate_art, degenerate_art_woke}. Echoing this sensibility, some of the styles and artists from this exhibition, like Picasso, were among the lowest rated in WikiArt. 

Realist biases can also limit the creative potential of visual generative AI. A throughline in HCI research on visual generative AI models is that artists often enjoy the glitches or surrealism of text-to-image (T2I) models \cite{shelby2024generative_ai, taylor2025straightening, chang2023prompt}, leading visual artists to critique the increasing realism of newer T2I models \cite{taylor2025straightening}. It follows that, aesthetic evaluation models that privilege realism may poorly align with the taste of some visual artists.

So far, we discussed the harms of being \textit{excluded} by LAP's algorithmic gaze. However, realism biases can also lead to harm through non-consensual \textit{inclusion}. We found that the LAION-Aesthetics 6.5+ Dataset is disproportionally made of images from independent visual artists and photographers. Not only have these communities lost jobs due to visual generative AI \cite{Merchant_2025_ai_jobloss, jiang2026professional}, but this suggests that their art may be driving the development of these technologies. As increasing photorealism has the potential to exacerbate the greatest harms of visual generative AI—like deepfakes \cite{mink2024deepfake}, job replacement \cite{jiang2023ai}, and sexual abuse imagery \cite{abdulla2025deepfakes}—we encourage researchers to reconsider the value placed on realism in visual evaluation.

\subsubsection{The Male Gaze}

We found that images with captions mentioning women were more likely to included in the LAION-Aesthetics 6.5+ Dataset, while those mentioning men or LGBTQ+ people were more likely to be excluded. It follows that datasets curated and models evaluated with LAP may also reflect these biases. In doing so, LAP reinforces the heterosexual male gaze in western visual art history \cite{mulvey_male_gaze_1975, berger2008ways} at scale. According to \citet{berger2008ways}, \textit{``Women [in western art] are depicted in a quite different way from men — not because the feminine is different from the masculine — but because the ideal spectator is always assumed to be male and the image of the woman is designed to flatter him.''} We found that LAP was designed according to the taste of one man: LAION's founder.

The male gaze is not new to computer vision. In 1973, male researchers at the University of Southern California digitized an image of the \textit{Playboy} model Lena Forsén to test image compression techniques \cite{mulvin2021proxies}. The ``Lena image'' became ubiquitous in computer vision research. Despite decades of criticism that the Lena image perpetuates misogyny \cite{chang2019brotopia, munson1996note}, IEEE publications did not prohibit its use until 2024 \cite{ieee_lena_ban_2024}. Much like Lena's image circulated without her consent among computer vision researchers for decades, images of (disproportionally) women now similarly circulate among computer vision researchers through the LAION-Aesthetics 6.5+ Dataset.

Although AI fairness researchers often critique data exclusions, \citet{hoffmann2021terms} critiques including marginalized communities in oppressive technologies. Recently, women have been immensely harmed by the proliferation of visual generative AI models. The United Nations has warned that generative AI is a major contributor to violence against women \cite{un_violence_women}. This violence has been facilitated by a rise of commercial AI-based nudification apps, often specifically designed to create nude images of women \cite{gibson2025analyzing}. However, this problem is not unique to dedicated nudification apps. At the time of our writing, thousands of men on X (formerly Twitter) have began using Grok to create and share NCII and CSAM of women and girls \cite{grok_ncii}. Of course, we cannot know whether these proprietary models rely on the LAION-Aesthetics Predictor or Datasets. Nevertheless, we expect that aesthetic evaluation models that increase the likelihood of women being included in AI training data risks exacerbating these harms. Much like recent work on how concept filtering impacts T2I CSAM generation \cite{cretu2025csam_laion_filtering}, we encourage future research on the relationship between aesthetic filtering and the generation of sexual imagery.

\subsection{Reevaluating Evaluation}

\subsubsection{Pluralistic Aesthetic Evaluation} 

We found numerous aesthetic biases in the behavior and design of LAP. However, in raising these issues, we do not advocate for designing ``better'' universal aesthetic evaluation models or datasets. As assessing the visual aesthetic quality of an image is deeply subjective and value-laden \cite{bourdieu1984distinction}, any attempt to represent the aesthetic quality of an image with a single number will be deeply flawed. As an alternative to this normative problem formulation, we encourage researchers to make use of more descriptive evaluating metrics \cite{rottger2022two_paradigms}. Although LAP is often used as a general measure of aesthetic quality \cite{ding2024understanding_data_poison, laion_eval_1, gao2024fbsdiff, he2023learning}, we found that it could be more accurately described as a measure of photorealism. As described above, there are a number of limitations to using realism to assess the quality of an image. However, a model that measures photorealism is not in-and-of-itself harmful. Rather, we caution researchers against treating photorealism as a prescriptive rather than descriptive property of images, conflating how images \textit{ought} to be with merely how images \textit{can} be.

To support more pluralistic AI alignment \cite{sorensen2024roadmap_plural}, we encourage the use of descriptive measurements when evaluating visual generative AI model outputs and training data. Abandoning the flawed construct of universal ``aesthetics'' would allow researchers to more clearly articulate \textit{which} aesthetic values their visual generative AI models prioritize, such as photorealism or particular visual cultures. For instance, Tencent's Hunyuan-DiT T2I model is specifically designed to improve Chinese visual element understanding \cite{li2024hunyuan}. In addition to improving transparency, more descriptive evaluations may help artists choose models that better align with their tastes, perhaps addressing artists' critique of increasing T2I photorealism \cite{shelby2024generative_ai, taylor2025straightening}. The seeds of more descriptive aesthetic evaluation can actually be seen in the 2012 AVA dataset itself \cite{murray2012ava}. Recognizing the limits of de-contextualized aesthetic evaluation, Murray et al. grouped similar photography challenges to measure 14 different photographic styles, such as ``motion blur'' and ``long exposure.''

The companies developing T2I models may have little incentive to shift from prescriptive to descriptive aesthetic evaluations. For-profit companies go to great lengths arguing that their models supplant competitors' models. For instance, Meta researchers manipulated benchmarks to give an inflated impression of Llama 4's performance \cite{meta_cheating}. Acknowledging that different T2I models may be better in different contexts undermines companies' efforts to prove that \textit{their} models are uniquely ``state-of-the-art.'' Much like corporate AI rhetoric, academic AI research is increasingly driven by benchmark competition \cite{gururaja2023benchmark_paradigm, orr2024benchmark_sport}, often without questioning the assumptions embedded in said benchmarks \cite{raji2021benchmark}. We encourage future researchers to examine what the universal aesthetics measurements used by companies and AI researchers \textit{actually} measure.

\subsubsection{Combining an Audit and Trace Ethnography}

In this work, we were able to examine upstream design decisions contributing to downstream biases by combining an audit and trace ethnography. Auditing is a core FAccT methodology, with the regex ``audit*'' appearing in nearly half of all FAccT publications. With some notable exceptions \cite{miceli2021documenting, marda2020data, riley2024overriding, scheuerman2024walled, passi2019problem, baack2024critical}, ethnography is far less common at FAccT. As \citet{seaver2017algorithms} argues, audits can help identify the existence of algorithmic biases while ethnographies can help explain where these biases may have originated by focusing on the people behind the algorithm. In addition to focusing on the harms of AI models themselves, responsible AI researchers are increasingly advocating for ``studying up'' the powerful people behind AI development and deployment \cite{barabas2020studying, kawakami2024studying}. In service of these goals, we advocate for combining audits and ethnographies as a socio-technical method for evaluating algorithmic systems.

Trace ethnographies can be a powerful tool for studying AI development. While FAccT researchers do not need permission from developers to audit an algorithmic systems, one would likely need permission to conduct participant observations \cite{scheuerman2024walled}. For example, prior ethnographic FAccT research has relied on access to field sites like tech companies \cite{scheuerman2024walled, passi2019problem} or a police headquarters \cite{marda2020data}. However, maintaining access can limit critical scholarship \cite{scheuerman2024walled}. Even in the case of ``open'' models, development practices may be opaque \cite{widder2024open}. We were only able to understand out how LAP was made by finding an explainer video in an old Discord message. Herein lies a lesson for future ethnographic FAccT research. As \citet{seaver2017algorithms} suggests, ethnographies of algorithms may require ``scavenging''  for diverse materials, including formal documentation, datasets, academic papers, and social media conversations. Nevertheless, this scavenging may pose ethical challenges. For example, studying an unlisted YouTube video—even if posted in a public Discord—could raise ethical concerns. We have chosen to include this video in our analysis due to LAION's advocacy for data scraping and refusal to remove artists' work from their datasets \cite{Robert_Kneschke_v_LAION}. That said, the ``public-ness'' of data does not always make its analysis ethical \cite{zimmer2010but}.

Finally, we caution that trace ethnographies are no substitute for more traditional ethnographic approaches, like extended participant observation and interviews. In our study, for instance, questions remain as to why the founder of LAION chose the datasets from which he built LAP. Likewise, it is not clear how members of LAION make sense of the widespread circulation of LAP. While our trace ethnography presents an account of how LAP was developed, extended participant observation would likely be needed to understand the culture of LAION within which LAP is situated. While we see potential for trace ethnographies to complement audits in FAccT research by embodying AI developers, this is not the only potential contribution of ethnographic analyses. Similar to \citet{forsythe2001studying}'s research on how expert system developers conceptualize ``experts,'' for example, we call for future ethnographic research into the how aesthetic evaluation researchers conceptualize ``aesthetics.''

\section{Conclusion}

Generative AI models and algorithmic systems are increasingly mediating our media, art, and culture. Therefore, it is imperative to understand for \textit{whose taste} these systems are designed. In this work, we audited the LAION-Aesthetics Predictor (LAP) model, finding that LAP reflects imperial, realist, and male aesthetic sensibilities. Through a trace ethnography of public materials, we found that the development of LAP reflects similar biases. In light of these findings, we call for greater attention to aesthetic biases as a source of representational harm and advocate for more pluralistic evaluations. Much like we examine the algorithmic gaze of LAP, we encourage future research into the cultural consequences of algorithmic ``quality'' assessment.

\section{Generative AI Usage Statement}

No generative AI tools were used to write this manuscript.

\begin{acks}
We thank Amy Bruckman, Rose Chang, Mary Gray, Sireesh Gururaja, Anna Fang, Oliver Haimson, Shivani Kapania, Franchesca Spektor, Cella Sum, David Gray Widder, Sherry Tongshuang Wu and our reviewers for their feedback on this work. We also thank the Science and Technology Studies Working Group at Carnegie Mellon University for providing early feedback. This work was supported, in part, by the National Science Foundation (award number 2442153).
\end{acks}

\bibliographystyle{ACM-Reference-Format}
\bibliography{refs}

\end{document}